\documentclass[seceq]{jjap2}
\title{Controlling the Intrinsic Josephson Junction Number in a $\mathbf{Bi_2Sr_2CaCu_2O_{8+\delta}}$ Mesa}

\author{\textsc{Li-Xing You}$^{1,2,}$\thanks{E-mail address: lixing@mc2.chalmers.se}, \textsc{Pei-Heng Wu}$^{1}$, \textsc{Wei-Wei Xu}$^{1}$, \textsc{Zheng-Ming Ji}$^{1}$ and \textsc{Lin Kang}$^{1}$}

\inst{$^{1}$ Research Institute of Superconductor Electronics(RISE),
Department of Electronic Science and Engineering, Nanjing University, Nanjing 210093, China \\
$^{2}$ Quantum Device Physics Laboratory, Department of
Microtechnology and Nanoscience, Chalmers University of
Technology, SE-412 96 Goteborg, Sweden \\}

\abst{ \indent In fabricating
$\mathrm{Bi_2Sr_2CaCu_2O_{8+\delta}}$ intrinsic Josephson
junctions in 4-terminal mesa structures, we modify the
conventional fabrication process by markedly reducing the etching
rates of argon ion milling. As a result, the junction number in a
stack can be controlled quite satisfactorily as long as we
carefully adjust those factors such as the etching time and the
thickness of the evaporated layers. The error in the junction
number is within $\pm 1$. By additional ion etching if necessary,
we can controllably decrease the junction number to a rather small
value, and even a single intrinsic Josephson junction can be
produced.}

\kword{intrinsic Josephson junctions, junction number, argon ion
etching, etching rates, $I-V$ curves}

\begin{document}
\maketitle

\section{Introduction}

Since the discovery of intrinsic Josephson junctions (IJJs) in
1992 \cite{Kleiner:PRL92}, considerable attention has been
attracted and much interest aroused, not only for the rich
treasury of nonlinear dynamics of IJJs but also for their possible
applications at high frequencies such as terahertz oscillators
with high power output and quantum voltage standard. With the
strongest anisotropy among high-temperature superconductors,
$\mathrm{Bi_2Sr_2CaCu_2O_{8+\delta}}$ (BSCCO) has been proven to
be the best material for fabricating IJJs. One of the main
challenges in the field is to fabricate such junction stacks in
which their number of junctions can be very well controlled.
Although the successful fabrication of IJJs containing only a few
junctions has been reported, the precise control of junction
number (particularly, in making a single intrinsic Josephson
junction) seemed difficult.
\cite{Yurgens:PRB96,Yurgens:APL97,Wang:APL01,Irie:PhysicaC02,Odagawa:IEEE99,Doh:PRB00}.

\par
Recently, we have modified the conventional fabrication process of
IJJs by reducing the etching rates of BSCCO and of the covering
metal layers during the Argon ion etching process. In a
three-terminal mesa we can control the junction number with an
error of $\pm 1$ \cite{You:SUST03}. However, the effects of the
surface junction and the layer damaged by the ion etching can
always be seen in the measured $I-V$ curves. The former prevents
us from obtaining consistently the value of the supercurrent of
the first branch, while the latter makes it impossible to reduce
the junction number in the stack to below two (the damaged layer
is approximately 2.7 nm thick). In this article, we report the
fabrication method of four-terminal mesas with further
improvement. With reduced etching rates and additional ion etching
process, we are now able to control the number of junctions in a
stack to the accuracy of one junction in each run. In other words,
we can controllably decrease the number of junctions in a stack to
a rather small value, and even single intrinsic Josephson junction
can be obtained.

\section{Experimental and Discussion}

\begin{figure}[tb]
\begin{center}
\includegraphics[width=.6\textwidth]{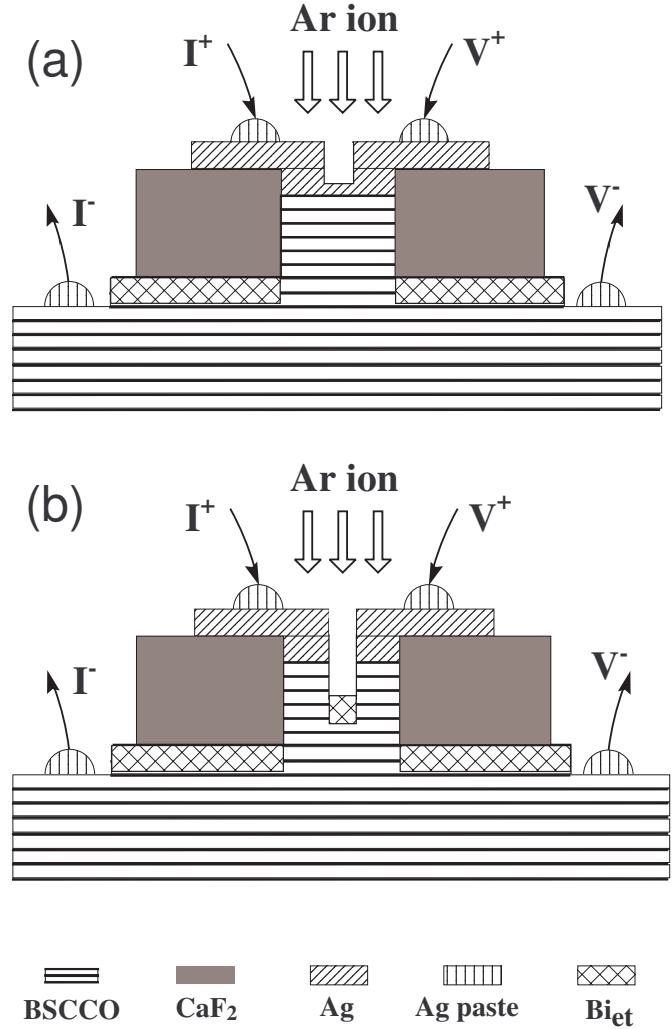}
\end{center}
\caption{Schematic view of IJJs of mesa structure. (a) when $T_2
\leq (d_1+d_2)/ER_{\mathrm{Ag}}$; (b) when $T_2 >
(d_1+d_2)/ER_{\mathrm{Ag}}$, where $\mathrm{Bi_{et}}$ represents
the BSCCO layer damaged during the ion milling.} \label{4-probe}
\end{figure}

BSCCO single crystals with $T_c$ of approximately 90 K are grown
using a traveling solvent floating zone (TSFZ) method
\cite{Lin:PhysicaC00}. After a small slice (typically of the size
of $\mathrm{2\ mm \times 2\ mm \times 0.1\ mm}$) is cleaved from a
bulk single crystal, a silver layer $d_1$ thick is evaporated
promptly onto the its fresh surface. Then the BSCCO slice is fixed
on a Si substrate for further fabrication. With the photoresist
sprayed, a square of $\mathrm{16 \times 16\ \mu m^2}$ is
photolithographically patterned onto the sample. Argon ion milling
is then carried out for a time duration of $T_1$ to make a mesa of
BSCCO which is promptly isolated by an evaporated layer of
$\mathrm{CaF_2}$. By ultrasonic rinsing in acetone, the top of the
mesa becomes free of $\mathrm{CaF_2}$ and the photoresist in a
lift-off process. Subsequently, a second silver layer $d_2$ thick
is evaporated onto the sample. Two separate electrodes,
$\mathrm{4\ \mu m}$ apart and each of the sizes of $\mathrm{6
\times 16\ \mu m^2}$, are photolithographically patterned and
electric leads can be glued onto them using the silver paste. For
four-terminal measurements, two more electrodes can be easily
formed on the base of the BSCCO slice directly. The schematic
diagram is shown in Fig. \ref{4-probe}.

\par

The above-mentioned top electrodes are made by argon ion milling
for a time duration of $T_2$. In doing so, obviously, not only the
top Ag layer ($d_2$ thick), but also the one under it ($d_1$
thick) and the BCCCO sample further down will probably be etched
as well, depending on the etching time, etching rate, and the
thickness of the silver layers. Supposing that the etching rates
for Ag and BSCCO are $ER_{\mathrm{Ag}}$ and $ER_{\mathrm{Bi}}$,
respectively, the junction number in the resulting stack can be
estimated to be $(T_1-d_1/ER_{\mathrm{Ag}})\cdot
ER_{\mathrm{Bi}}/d_0$ (for $T_2 \leq (d_1+d_2)/ER_{\mathrm{Ag}})$
or $(T_1-T_2+d_2/ER_{\mathrm{Ag}}) \cdot ER_{\mathrm{Bi}}/d_0$
(for $T_2 > (d_1+d_2)/ER_{\mathrm{Ag}})$, respectively, where
$d_0$ is the distance between two adjacent superconducting layers
along the $c$ axis (typically 1.54 nm for BSCCO). Furthermore, our
experiments have indicated that the base of BSCCO seems to be
damaged during the ion millings, making the stack slightly taller
than its geometric height. In other words, an empirical fitting
parameter $d_{\mathrm{et}}$ should be included to correctly
estimate the junction number in a stack. Thus the junction number
$N$ is
\begin{equation} \label{eq1}
N=[(T_1 - d_1/ER_{\mathrm{Ag}}) \cdot ER_{\mathrm{Bi}}
+d_{\mathrm{et}}]/d_0 \quad for \quad T_2 \leq
(d_1+d_2)/ER_{\mathrm{Ag}}
\end{equation}
or
\begin{equation} \label{eq2}
 N=(T_1-T_2+d_2/ER_{\mathrm{Ag}}) \cdot ER_{\mathrm{Bi}}/d_0 \quad for \quad T_2 > (d_1+d_2)/ER_{\mathrm{Ag}}
\end{equation}
We note that in the fabrication $T_1$ is always larger than
$d_1/ER_{\mathrm{Ag}}$, thus in the structure described by eq.
(\ref{eq1}) the minimum junction number is determined by
$d_{\mathrm{et}}/d_0$, denoting that we cannot obtain a stack with
less than $d_{\mathrm{et}}/d_0$ junctions in it. While in the
structure described by eq. {\ref{eq2}}, there is no such a minimal
limit to the junction number, because that $T_1$ and $T_2$ can be
adjusted independently, and for sufficiently long $T_2$ we can
certainly break the bound set by $d_{\mathrm{et}}$.

\par

In the ordinary fabrication of IJJs, ion milling is usually
carried out at high energy, resulting in high etching rates of
both Ag and BSCCO. The typical values were 76 nm/min for Ag and 18
nm/min for BSCCO in our previous work \cite{You:SUST02,
You:CSB03}. At such high etching rates, it seemed difficult to
precisely control the number of junctions in a stack by
controlling $T_1$ and $d_1$. Besides, to ignore the layer damaged
by the ion etching also causes an error in the junction number. In
our previous work, the error of the junction number is $\pm 5$.
The basic idea for the present work is to reduce the rates
markedly so that precise control of the junction number becomes
feasible. By adopting a neutral ion energy of 250 eV and a flow
density of 0.38\ mA/cm$^2$ during the argon ion milling, we lower
the etching rates to 9.3 nm/min for Ag and 1.3 nm/min for BSCCO.
Accordingly, the empirical fitting parameter $d_{\mathrm{et}}$ is
2.7 nm, which is obtained from the statistics of many samples
\cite{You:SUST03}.

\par

\begin{figure}[tb]
\begin{center}
\includegraphics[width=.6\textwidth]{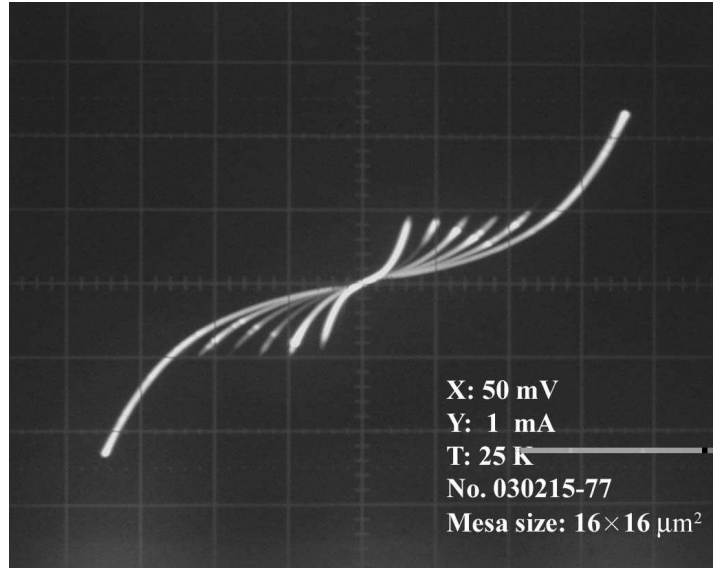}
\end{center}
\caption{$I-V$ characteristics of a stack with 6 junctions
(including a surface junction).} \label{IV1}
\end{figure}

The measurements are carried out in a cryocooler with samples kept
at typically 25 K. For one of our samples ($d_1$=68 nm, $d_2$=102
nm, $T_1$=12 min, $T_2$=18 min), the number of junctions in the
mesa is approximately 5 as given by eq. (\ref{eq1}). Fig.
\ref{IV1} shows the $I-V$ curves of the sample. Judging from the
quasi-particle branches, there are indeed 5 junctions in the
stack, just the same as what eq. (\ref{eq1}) gives. However, at
the interface between the BSCCO mesa and the Ag layer, there
exists a surface junction; taking this into account makes a total
junction number of 6. Then the error in N is 1. For the more than
30 samples that we have fabricated, the error of N is within
$\pm1$ for junction numbers from 3 to 12; this indicates that by
controlling $T_1$, $d_1$ and the etching rates, the junction
number in a stack can really be controlled quite satisfactorily,
as shown by eq. (\ref{eq1}).

\par

Due to the existence of a surface junction, the supercurrent is
not registered in the $I-V$ curves in Fig. 2, which was discussed
in detail by Doh et al. \cite{Doh:PRB00} In fact, surface
degradation reduces the critical temperature of the surface
junction; and when the experimental temperature is above it, the
junction will manifest itself as a nonlinear resistor, thus
quenching the supercurrent in IJJs. If we increase the etching
time to $T_2 > (d_1+d_2)/ER_{\mathrm{Ag}}$, the surface junction
will be broken into two parts [Fig. \ref{4-probe}(b)], and their
resistances will be included in those of the top electrodes
accordingly; thus in four-terminal measurements the influence of
the surface junction on $I-V$ curves will be eliminated. With
further etching, the structure becomes U-shaped; under general
conditions, the right and left stacks are of equal sizes and act
as leads for the measurements, and only the bottom stack is of
importance in what follows. The effects of the sizes of the right
and left stacks or the effects of their irregularities will be
discussed elsewhere.

\par

\begin{figure}[tb]
\begin{center}
\includegraphics[width=.6\textwidth]{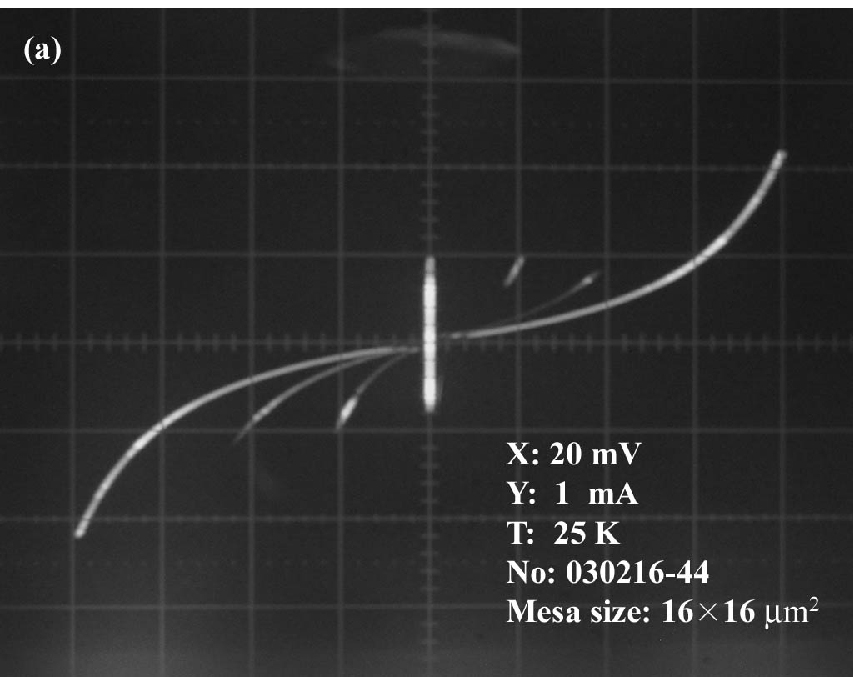}
\includegraphics[width=.6\textwidth]{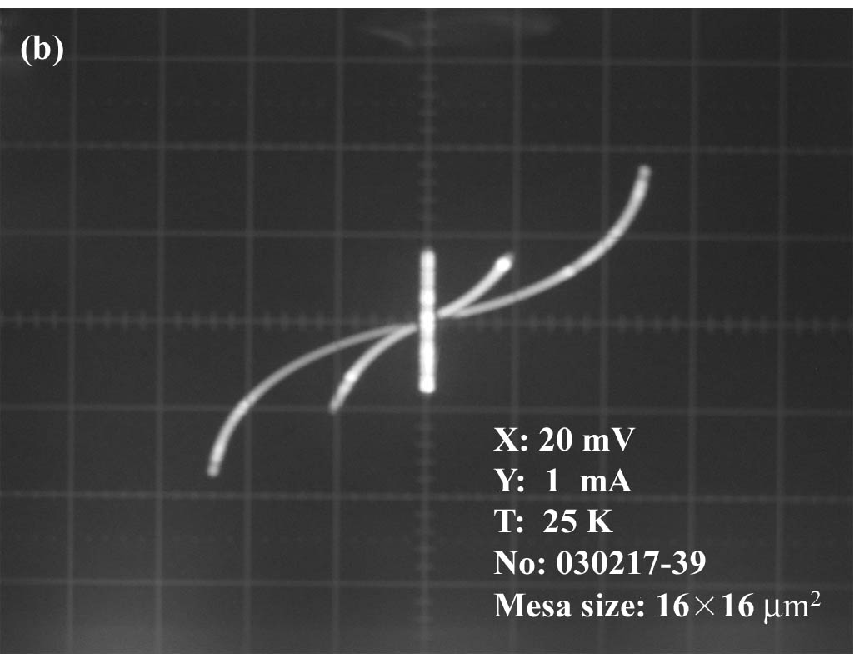}
\includegraphics[width=.6\textwidth]{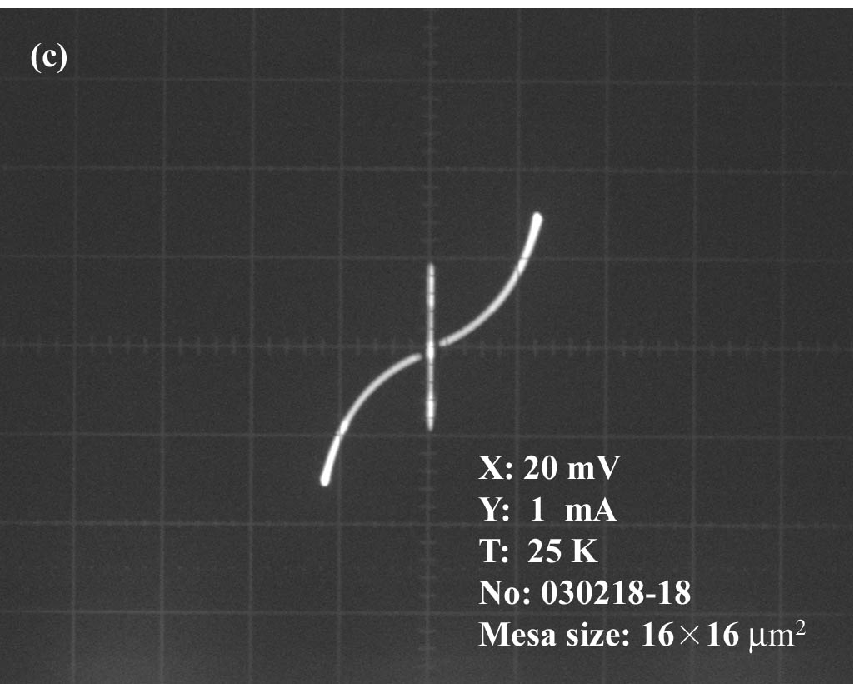}
\end{center}
\caption{Current-voltage characteristics of IJJs. Successive
etchings reduce the junction number in a stack from (a) 3, to (b)
2, and to (c) 1. Each etching lasts one minute.} \label{IV2}
\end{figure}

When $d_1$=68 nm, $T_1$=12 min, and $d_2$ =102 nm, $T_2$ = 19 min,
we obtain the $I-V$ curves of 3 junctions as shown in Fig.
\ref{IV2}(a). The junction number in the curves agrees well with
the result given by Eq. \ref{eq2}. If, with this same sample, we
increase $T_2$ by 1 min in each of the following etchings, the
junction number in the stack will be reduced to 2 and 1
accordingly as shown by the $I-V$ curves in Fig. \ref{IV2}(b) and
Fig. \ref{IV2}(c). The $I-V$ curves in Figs. \ref{IV2}(a) to (c)
indicate that, apart from the reduction in junction number, there
is no appreciable change in the shape of the $I-V$ curves nor in
the values of the supercurrent; i.e., no degradation is observed
after several etchings and thermal cycles.

\par
From Eq. \ref{eq2}, we know that the errors of the thickness of
the metal layers, the etching rates and the etching time lead to
the error of the junction number. The results above show that we
can precisely control the junction number of IJJs by additional
ion etching process, thus successfully eliminate the error $\pm 1$
in the junction number. If we want a sample of IJJs containing $n$
junctions, we may set the parameters ($T_1$, $T_2$, $d_2$) in eq.
(\ref{eq2}) to satisfy $N=n+1$. The obtained junction number $N_0$
should be between $n$ and $n+2$ because of the existence of error
$\pm 1$. Optional additional ion etching process can be adopted to
reduce the junction number to $n$ with the accuracy of one
junction in each run. This method is so successful that by using
it we have successfully fabricated several samples with only one
intrinsic Josephson junction of the sizes of $\mathrm{16 \times
16\ \mu m^2}$ or $\mathrm{8 \times 8\ \mu m^2}$.

\section{Conclusions}

In order to control the junction number in a stack, we can adjust
such factors such as the etching time, the etching rates and the
thickness of the covering Ag layers. In determining the junction
number, in principle, they should have equal weight; but in
practice, if we want to keep the etching time or the layer
thickness within reasonable ranges, a low etching rate is of
extreme importance. By markedly reducing the etching rates and
carrying out optional additional ion etching processes, we have
improved the conventional fabrication process of IJJ in
four-terminal structures; thus we are now able to obtain IJJs
stacks containing any number of junctions down to only one. This
method provides us with the possibility of looking at how the
layer number in a IJJs stack affects its dynamical behavior or its
superconductivity. Also under way are the studies on the
properties of one intrinsic Josephson junction singled out from a
stack.

\section*{Acknowledgment}

We would like to thank Y. Cao and W. X. Cai for their help during
the measurements. This work was supported by a grant from the
Major State Basic Research Development Program of China (No.
G19990646) and the National High-tech Research and Development
Program of China.

\end{document}